\documentclass[10pt, conference, letterpaper]{IEEEtran}

\IEEEoverridecommandlockouts
\usepackage{cite}
\usepackage{amsmath,amssymb,amsfonts}
\usepackage{algorithmic}
\usepackage{graphicx}
\usepackage{textcomp}
\usepackage{xcolor}
\usepackage{balance}
\def\BibTeX{{\rm B\kern-.05em{\sc i\kern-.025em b}\kern-.08em
    T\kern-.1667em\lower.7ex\hbox{E}\kern-.125emX}}
\begin{document}

\title{A novel Two-Factor HoneyToken  Authentication Mechanism 
}

\author{\IEEEauthorblockN{Vassilis Papaspirou}
\IEEEauthorblockA{\textit{Department of Computer Science} \\
\textit{University of Thessaly}\\
Volos, Greece \\
vpapaspyrou@uth.gr}
\and
\IEEEauthorblockN{Leandros Maglaras}
\IEEEauthorblockA{\textit{Cyber Technology Institute} \\
\textit{De Montfort University}\\
Leicester, UK \\
leandros.maglaras@dmu.ac.uk}
\and
\IEEEauthorblockN{Mohamed Amine Ferrag}
\IEEEauthorblockA{\textit{Department of Computer Science} \\
\textit{University of Guelma}\\
Guelma, Algeria \\
ferrag.mohamedamine@univ-guelma.dz }
\and
\IEEEauthorblockN{Ioanna Kantzavelou}
\IEEEauthorblockA{\textit{Department of Informatics} \\
\textit{University of West of Attica}\\
Athens, Greece \\
ikantz@uniwa.gr}
\and
\IEEEauthorblockN{Helge Janicke}
\IEEEauthorblockA{\textit{Cyber Security Cooperative Research Centre} \\
\textit{Edith Cowan University}\\
Australia \\
helge.janicke@cybersecuritycrc.org.au}
\and
\IEEEauthorblockN{Christos Douligeris}
\IEEEauthorblockA{\textit{Department of Informatics} \\
\textit{University of Piraeus}\\
cdoulig@unipi.gr }
}

\maketitle

\begin{abstract}

The majority of systems rely on user authentication on passwords, but passwords have so many weaknesses and widespread use that easily raise significant security concerns, regardless of their encrypted form. Users hold the same password for different accounts, administrators never check password files for flaws that might lead to a successful cracking, and the lack of a tight security policy regarding regular password replacement are a few problems that need to be addressed.
The proposed research work aims at enhancing this security mechanism, prevent penetrations, password theft, and attempted break-ins towards securing computing systems. The selected solution approach is two-folded; it implements a two-factor authentication scheme to prevent unauthorized access, accompanied by Honeyword principles to detect corrupted or stolen tokens. Both can be integrated into any platform or web application with the use of QR codes and a mobile phone.


\end{abstract}


\section{Introduction}

 In all areas, such as banks, government applications, the pharmaceutical
sector, military organisations, educational establishments, etc., security issues are growing today. Government institutions set guidelines, pass regulations, and compel organizations and
agencies to conform with these standards, with wide-ranging implications of non-compliance.
In these various and varied industries with a common weak link being passwords, there are many challenges when it comes to security issues. To verify the identity of the user,
most applications today rely on static passwords. These keys, though, come with serious security issues for administrators. Users prefer to use easy-to-guess passwords, use different accounts with the same password, write passwords or save them on their computers unencrypted. Moreover, although dedicated systems, called password managers, can offer secure password storage and retrieval,  only a small fraction of users use them \cite{zhang2019people}. In addition, hackers have the choice of using many password stealing methods, such as shoulder surfing, snooping, sniffing, guessing, etc. Several best practices have been suggested for the use of passwords.
Some of them are very difficult to use and
others do not fulfill the security needs of the organization.
To overcome the password problem, two factor authentication using devices such as tokens and
ATM cards has been suggested and has been shown to be difficult to hack  \cite{twoFactor}. There are several limitations of two-factor authentication, including the
cost of purchasing, issuing, and handling tokens or cards.
From the point of view of the user, having more than one two-factor authentication
methods demands the purchase of several tokens/cards that are likely to be misplaced or stolen.   

 Traditionally, cell phones have been considered
a device for making phone calls.
But today, the use of cell phones has been generalized to send
calls, review addresses, shop contacts, etc., provided the developments in hardware and software. Also, opportunities for smartphone access have expanded.
Cell phones combine infra-red, Bluetooth, 3G,
and WLAN connectivity, on top of normal GSM connectivity.
For contact purposes, most of us, if not all of us, hold cell phones. 
Several accessible mobile banking services take advantage of mobile computer enhancement capabilities. From the ability to collect account balance information in the form of SMS messages to the use of WAP and
Java along with GPRS to allow fund transfers between accounts, stock trading, and direct payment confirmation through the phone's micro browser.

The principle of using passwords and smart cards to authenticate customers is an old idea going back 40 years now. Since then many systems with two-factor authentication mechanisms were developed. However since the smart card may be intercepted and the data contained in the smart card may be duplicated, the reliability of two-factor authentication may be breached, and the number of potential passwords can be limited and users could forget or lose their passwords. 

Biometric authentication was adopted to authenticate users by using their biometric characteristics due to those issues.
Scholars have suggested biometric authentication system since back in 1999 which enhances some facets of two-factor authentication since biometric features have greater entropy and can not be missed and are rarely lost. One drawback, though is that biometric characteristics are not entirely confidential since one can "steal" biometric characteristics from others for example, the fingerprint can be retrieved from a mug used by the suspect and the facial features can be obtained from an image of a user. Combining all these three variables together is a way to mitigate these concerns. This technique is often referred to as three-factor authentication, and has been greatly adapted by cloud-based applications.\cite{yu2014efficient}

SIM cards are available in varying storage sizes. Related memory utilization of the SIM card connected with it plays a part in deciding the effectiveness of cloning the SIM card, more memory stored on the original SIM card than the longer the Ki A8 algorithm cracking process on the SIM card. Problems resulting from the above perspective relating to the inclusion of the A8 algorithm inserted in any SIM card used by telecommunications users to duplicate or replicate the SIM card are detrimental to the privacy and protection of cell phone users on either side.
The purpose of the SIM card cloning research is to provide an alert to consumer safety and provide a dedicated SIM card to tackle SIM card cloning criminal investigations along with their abuse of data.Subscriber Authentication Based on IMSI (Stored on SIM) and Andom Number Generator/RAND (Provided by Network), SIM card cloning authentication will be further investigated by comparing the network login response of the customer to the mobile service network. The Random Number Generator (RAND) includes an algorithm A3 (Provided by Network) such that RAND participates in the process of cloning the SIM card in order to adapt the algorithms contained in the SIM card A8 to A3 algorithms contained in the user data of the connected network authentication. \cite{anwar2016forensic}

Scholars have already demonstrated that by launching a cross-platform infection attack, an attacker is able to compromise another device, either a PC or a cell phone. Prototypes of proof-of-concept demonstrate that such attacks are feasible and thus it is not fair to preclude them from the mobile 2FA scheme adversary model. The intruder will snatch all authentication tokens and impersonate the rightful user when both 2FA devices are infected, regardless of what individual smartphone 2FA instantiation is used.We carry out attacks against various instantiations of mobile 2FA schemes implemented by banks and common Internet service providers to help our argument.

Schemes with 2FA OTPs created on the client side, such as Google Authenticator (GA), depend on pre-shared secrets. The configuration process of the GA app, used by hundreds of providers, including Google Mail, Facebook and Outlook.com, was evaluated. When the user allows GA-based authentication in his account settings, the GA initialization begins. A QR code is created by the service provider and shown to the user (on the PC) and scanned by the user's smartphone. All the information required to initialize GA with user-specific account details and pre-shared secrets is stored in the QR code. During the initialization process, scholars analysed the QR code submitted by Facebook and Google and defined the structure of the QR code. This includes information such as the scheme sort (counter-based vs. time-based), the service and account identifier, the counter (counter-based mode only the generated OTP duration and the mutual secret identifier.
In addition, all this material is provided in plain text. To check if GA supports any alternate initialization system, scholars \cite{dmitrienko2014security} reverse engineered the app with the JEB Decompiler and evaluated the internal apps. We have not found any alternate initialization routines, suggesting that this initialization protocol is used by all 32 service providers using GA.
The initialization message may be intercepted by a PC-residing malware (clear text encoded as an QR code). The attacker will then initialize its own version of the GA and can produce legitimate OTPs.

The use of 'honeywords' was introduced in order to detect whether or not the password file was stolen, i.e. a series of false passwords that are combined with the original password of the user and the hash values of these passwords (real passwords and honeywords) are contained in the password file. 
The adversary also does not know which one is the true password if this file is corrupted and all the hash values in the file are cracked. Note that LS identity and password are submitted by the customer or the adversary to request login.LS then checks if a password submitted is among the honeywords of a user, but even if this search succeeds, LS needs to review another protected component, HC, to see if the index of the honeyword retrieved corresponds to the actual password of the user. HC warns the administrator otherwise, as a honeyword signal has been detected that the password file might have been corrupted \cite{genc2017examination}.

Based on these findings and trying to combine the strengths of honeywords and 2FAs while at the same time keeping the system simple and easily integrated in any existing platform or system, we present in this paper a prototype of a novel security mechanism. We develop and propose an innovative security mechanism for web applications that produces both passwords and QR codes covering different login modes. The proposed system entitled "Two-Factor HoneyToken Authentication (2FHA)", combines the strengths of two-factor authentication and Honeyword technologies.  In the developed prototype a sms with 3 OTP passwords that correspond to 3 QR codes is sent to the user. Only one of these three elements is the correct token that can be used in order to continue.  This induces an extra layer of security adding more safety to the system. The proposed system offers enhanced security to the user while at the same time is simple and doesn't impose additional overhead during login.

The rest of the article is structured as follows. Section \ref{2FA} presents two-factor authentication principles and limitations. Section \ref{sec:honey} discusses honeywords principles. Section \ref{sec:prototype} presents the proposed system architecture and protopype and Section \ref{sec:concl} concludes the article and discusses future work.

\section{Two factor authentication}\label{2FA}

Two-factor authentication (2FA) is a security mechanism in which users use two separate  authentication keys to validate themselves, often referred to as two step verification or dual-factor authentication. This process is undertaken to help secure both the credentials of 
the user and the tools that can be used by the user. 
Two-factor authentication offers a higher degree of protection than one-factor authentication (SFA)-dependent authentication systems, in which the user only provides one factor, normally a password or passcode. Two-factor authentication strategies rely on a password-providing mechanism, as well as a second factor, typically 
either a safety token or a biometric factor, such as a fingerprint or facial scan.
Two-factor authentication brings to the authentication process an extra layer of security by making it more difficult for criminals to obtain access to computers or online accounts of an individual since it is not enough to know the victim's password alone to pass the authentication check. To monitor access to confidential applications and files, two-factor authentication has long been used and online service providers are gradually using 2FA to secure 
the identities of their customers from being used by hackers who have compromised a password database or used phishing campaigns to acquire user passwords\cite{twobirds}.

\subsection{What are authentication factors?}
There are many different ways in which more than one authentication mechanisms are used to authenticate anyone. 
Most authentication mechanisms usually rely on factors of information, such as a traditional password, 
whereas two-factor authentication methods incorporate either a possession factor or a factor of inherence \cite{ferrag2017authentication}.

Authentication factors, listed in approximate order of adoption for computing, include the following:
\begin{enumerate}
\item	A knowledge factor is when The user knows something, such as a password, a personal 
identification number (PIN) or some other sort of mutual secret.
\item	A possession factor is when a user has To accept authentication requests, the user has something, such as an ID card, 
a protection key, a cell phone, a mobile computer or a smartphone app.
 
\item	An inherence factor refers to anything intrinsic to the physical self of the 
individual is more generally considered a biometric element. 
This may be personal characteristics, such as fingerprints authenticated by a fingerprint scanner, are mapped to physical features. Facial and speech recognition are other widely used inherence variables. 
There are also the biometrics of behavior, such 
as keystroke dynamics, variations of gait or voice.

\item	A location factor typically denoted by the location from which an authentication attempt is made, can be implemented by restricting authentication attempts to specific devices in a specific location or more commonly, by monitoring the geographical source of an authentication attempt based on the Internet Protocol (IP) source address or some other geolocation detail, such as data from the Global Positioning System (GPS), 

\item	A time factor limits user authentication to a fixed time frame where it is allowed  to log in and limits access to the device beyond that window.
\end{enumerate}

It should be remembered that the vast majority of two-factor authentication mechanisms rely on the first three authentication factors, while multifactor authentication 
(MFA), which may rely on two or more separate passwords for more reliable authentication, can be used by systems that demand greater security.

\subsection{How does two-factor authentication work?}
In this section we briefly describe the process of a typical two factor authentication system \cite{ferrag2017authentication}.

\begin{itemize}
\item The user is asked by the program or by the website to log in.

\item The user enters what he or she knows—usually a username and password. Then a match is made by the site's server and the user is remembered.

\item The website creates a special authentication key for 
the user for processes that don't need passwords. 
The authentication function processes the key and 
it is checked by the site's server.

\item Then the site asks the user 
to start the second stage of login. 
While a variety of ways can be taken through this step, users must show that they only 
have what they will have, such as an identification key, ID card, smartphone or other mobile device. 
This is the 
factor for ownership.

\item During phase four, the 
user enters a one-time code created.

\item	The customer is authenticated and given access to the program or website after supplying all variables.
\end{itemize}

In technical terms, two authentication factors are required to obtain access to a device or facility at any point. 
Using two variables from the same group, though, would not constitute 2FA; for instance, it is always called SFA 
to require a password and a mutual secret since both belong to the same class of authentication factor: information. 
The user ID and password are not the 
most reliable as far as SFA services. 
One concern with password-based authentication is that generating 
and recalling good passwords requires awareness and diligence. 
Passwords need protection against many internal attacks, such as carelessly kept 
login credential sticky notes, old hard drives and vulnerabilities in social engineering. 
Passwords are often vulnerable to external threats, such as hackers using brute-force, dictionary or 
rainbow table attacks.

An intruder will typically break password-based protection mechanisms and steal corporate 
data, including personal information of users, provided ample time and money. 
Because of their low cost, ease of execution and 
familiarity, passwords have remained the most common type of SFA. 
Depending on how they are applied, several challenge-response questions can provide more security, 
and stand-alone biometric authentication approaches can also provide a more reliable SFA process.

\subsection{Types of two-factor authentication products}
There are several different 2FA deployment equipment and utilities — from tokens to radio frequency identification (RFID) cards to applications for smartphones \cite{ferrag2020authentication}.

It is possible to separate two-factor authentication devices into two categories: tokens that are provided to users to use 
while signing in and infrastructure or software that detects and authenticates entry for users who correctly use their tokens.

 Physical devices, such as key fobs or smart cards, may be authentication keys, or 
they may exist in applications like mobile or web apps that produce authentication PIN codes \cite{limbasiya2018advanced}. 
These authentication codes are normally created by a server, often known as one-time 
passwords (OTPs), and can be recognized by an authentication system or app as authentic. 
The authentication code is a short sequence connected to a specific computer, user 
or account that can be used once as part of an authentication process.
To accept, process and authorize — or reject — access to 
users who authenticate with their tokens, organisations need to install a framework. 
This may be implemented in the form of cloud applications, a dedicated hardware server, or supplied by a third-party provider as a service.

An significant feature of 2FA is ensuring that the authenticated user is granted 
access to all services the user is allowed for — and only those resources. 
As a consequence, one of 2FA's main functions is to 
connect the authentication method with the authentication data of an entity. 
Microsoft offers some of the required infrastructure for Windows 10 2FA service organisations through Windows 
Hello, and will work with Microsoft accounts, as well as authenticate users with Microsoft Active Dii.

\subsection{How 2FA hardware tokens work}

Hardware tokens for 2FA are 
available that support numerous authentication approaches \cite{reynolds2020empirical}. 
The YubiKey, a small Universal Serial Bus (USB) system that supports OTPs, public key encryption and authentication, 
and the Universal 2nd Factor (U2F) protocol developed by the FIDO Alliance, is a common hardware token. 
YubiKey tokens are sold by 
Palo Alto, California-based Yubico Inc.

When YubiKey users log in to an OTP-supported online site, such as Gmail, GitHub, or WordPress, they insert their 
YubiKey into their device's USB port, enter their password, select the YubiKey field, and then tap the YubiKey icon. 
YubiKey produces and inputs 
an OTP into the field. The OTP is a 44-character, single-use password; a special ID defining 
the authentication key associated with the account is the first 12 characters. 
The remaining 32 characters contain information that is encrypted using a key only known to the 
computer and the servers of Yubico that was generated during the initial registration of the account.
An OTP is submitted from an online 
service to Yubico for verification of authentication. 
The Yubico authentication server sends back a message verifying that this 
is the correct token for this user until the OTP is checked. Two authentication criteria have been given by the user: the information 
factor is the password, and the possession factor is the YubiKey.

\subsection{Two-factor authentication for mobile device authentication}

For 2FA, smartphones provide a number of possibilities, 
encouraging organizations to choose what suits best for them. 
A built-in camera can be used for face recognition or iris detection, and the 
microphone can be used for speech recognition. Certain applications are able to recognise fingerprints. 
GPS-equipped smartphones will check the 
location as an extra consideration. 
Also, Speech or Short Message Service (SMS) 
may be used as an out-of-band authentication channel.
For receiving authentication codes by text message or automatic 
phone call, a trustworthy phone number may be used. 
To participate in 2FA, a person needs 
to check at least one trustworthy phone number.
Both applications that support 2FA are available for Apple iOS, Google Android and Windows 10, 
allowing the phone itself to function as the physical interface to satisfy the ownership aspect. 
Duo Defense, headquartered in Ann Arbor, Mich., and acquired for \$2.35 billion by Cisco in 
2018, is a 2FA software provider whose solution allows 2FA consumers to use their trusted products. 
Before checking that the mobile device can still be trusted to 
authenticate the customer, Duo's platform first determines that a user is trusted.
The need to acquire an authentication code through text, 
voice call or email is replaced by authenticator apps. 
For example, users type in their username and password to access a 
website or web-based application that supports Google Authenticator — a knowledge factor. 
Users are then asked to 
type a number of six digits. 
Instead of having to wait a few seconds to answer 
a text message, an Authenticator produces the number for them. 
Every 30 seconds, these numbers alter 
and are different with every login.
Users complete the authentication process by entering the correct number 
and show custody of the correct unit — an ownership element. 

\subsection{Is two-factor authentication secure?}

There are several limitations of two-factor authentication, including the
cost of purchasing, issuing, and handling tokens or cards.
From the point of view of the user, having more than one two-factor authentication
method allows several tokens/cards to be held that are likely to be misplaced or stolen. 
Although two-factor authentication improves security—because access privileges are no longer dependent solely on  a password's strength,—two-factor authentication systems are just as reliable as their weakest part. 
Hardware tokens, for instance, depend on 
the security of the issuer or manufacturer. 
In 2011, when the technology firm RSA Security announced its SecurID authentication tokens had 
been stolen, one of the most high-profile examples of a compromised two-factor device occurred.
If it is used to circumvent two-factor authentication, the account recovery mechanism itself can often be subverted because it sometimes resets the 
existing password of a user and e-mails a new password to allow the user to log in again, bypassing the 2FA process. 
The corporate Gmail accounts of the chief 
executive of Cloudflare were compromised in this way.
Although 2FA is cheap, simple to implement and user-friendly based on SMS, it is vulnerable to multiple attacks. 
In its special publication 800-63-3, the National Institute of Standards and 
Technology (NIST) has discouraged the use of SMS in the 2FA services \cite{grassi2017draft}.
Due to cell phone number portability attacks, such as the Signaling System 7 hack, against the mobile phone network and malware, such as Eurograbber, that can be used to intercept or divert text messages, NIST concluded that OTPs sent via SMS are too vulnerable.From all the above factors the idea of 2HFA is created.

\section{Honeywords}\label{sec:honey}
The fundamental principle behind the Honeywords scheme is to adjust the password storage mechanism in such
a way that a password and a series of false passwords are associated with each account \cite{honeywords}.
The phony passwords are called honeywords. Sweetwords are
the union of both honeywords and the password.
As soon as the password is entered during the authentication
process, the password database is immediately detected to have been compromised. Therefore unlike traditional schemes, implementations focused on
honeywords can effectively detect violations of password databases.

\begin{figure}
    \centering
    \includegraphics{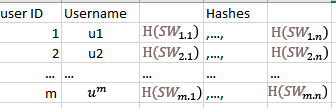}
    \caption{Credentials database of aLSin the Honey-words system}
    \label{fig:honey1}
\end{figure}

\begin{figure}
    \centering
    \includegraphics{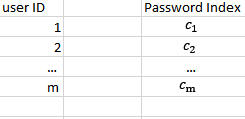}
    \caption{Data stored on aHC}
    \label{fig:honey2}
\end{figure}

The method of Honeyword is as follows.
During the authentication process, users select a username and a password, as with many traditional schemes. The
Login Server (LS) then produces honeywords for the password and maintains a record in the database of passwords. 
The ordering of the sweetwords is randomly
selected by the LS in each record.
In addition, LS sends the corresponding user ID and actual password index
to Honeychecker (HC), the auxiliary server built to store the password index.
Let ui and H() denote respectively the user name of user I and the
hash function used in the method. H(swi,j) denotes the hash of user i. jth sweetword. A standard example of a table
of qualifications is illustrated in Figure \ref{fig:honey1}.
HC saves the user IDs and the password index between the honeywords. During the authentication, no username or password itself is sent to HC. In comparison, HC is built as a hardened server that can only be reached by LS.
A standard structure of the
HC data is seen in Figure \ref{fig:honey2}.
Notice that only two kinds of messages
are accepted by HC: Check and Set 
To verify {\bf if j=ci, check(i, j)} implies that uf j=ci, HC returns True, otherwise False
is returned and a warning is activated.

The command set is structured as: 
{\bf Set (I j) indicates setting ci=j}. 
The user submits its username and password.LStries during the authentication process
to locate the corresponding record for that username in the credentials database.
If a record exists, LS computes the hash of the password
sent and attempts to find a match in the sweetword hashes. If no match occurs, then the password
sent is incorrect and access is refused.
LS sends the respective user ID and the
corresponding index to HC if there is a match.
First, HC seeks the record that fits the user ID and
compares the index value obtained with the one stored in its database. If the outcome is
valid, then access is provided.
Otherwise the HC returns incorrect, generates
an alert and notifies the administrators of the device policy.

Originally, the Honeywords scheme was constructed with the expectation that the opponent
could steal the hashed passwords and invert the hashes to obtain the passwords.
It is therefore presumed that both LS and HC will
not be abused by the attacker within the same time frame. The Honeywords mechanism defends passwords from brute-force and dictionary attacks mentioned in Section \ref{2FA}. The method attempts to prevent violations of the password database and seeks to prevent only offline dictionary attacks where the adversary is believed to have taken the hashes of the password and abandoned the system.

\section{The proposed Two-Factor HoneyToken  Authentication (2FHA) Mechanism}\label{sec:prototype}
%
In this article we introduce an alternative authentication method, for enhancing systems' security. The system combines two factor authentication with honeywords in order to make impossible for an attacker to bypass the authentication mechanism of the system. Even in the occasion that the attacker has access to the device that receives the token, e.g. by sim cloning, the proposed 2FHA method makes the authentication bypass unfeasible if not impossible. 

\begin{figure*}[h]
    \centering
    \includegraphics[width=0.85\linewidth]{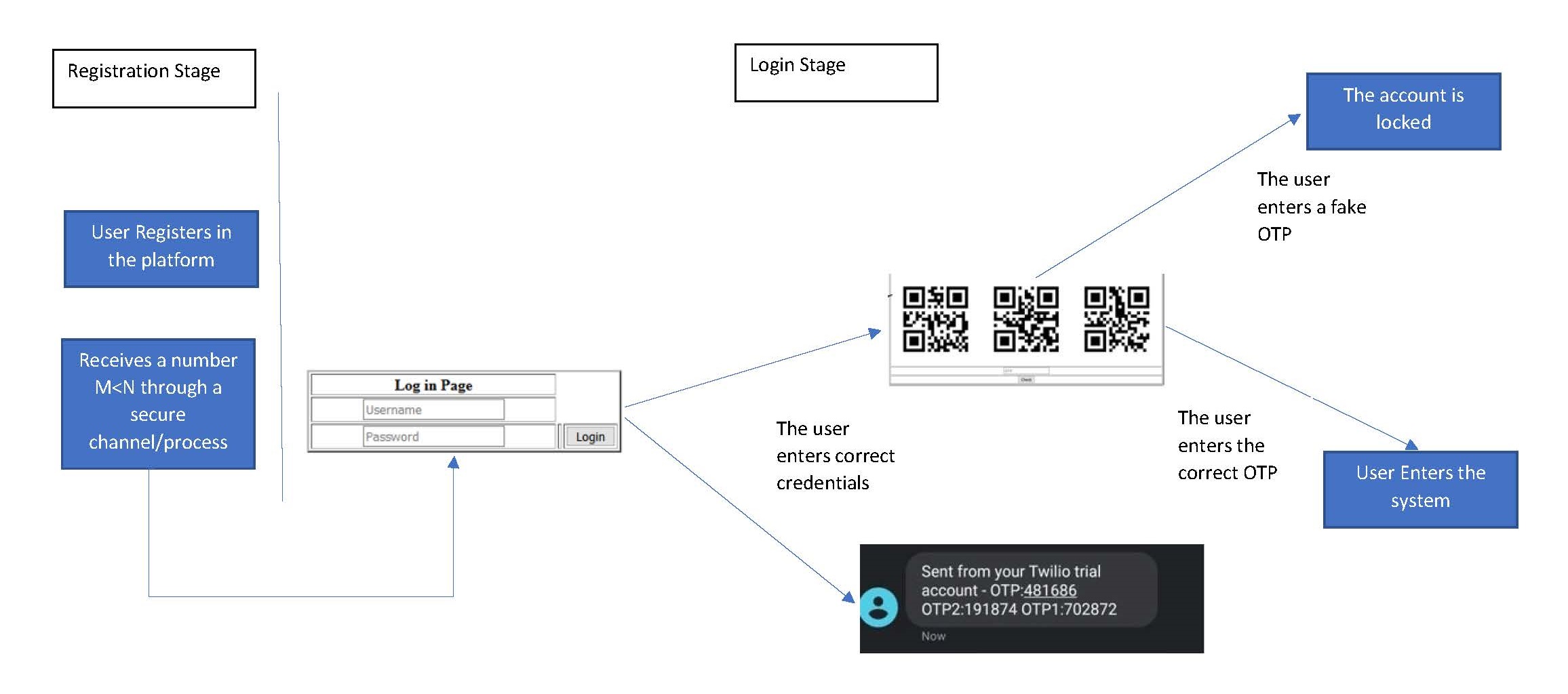}
    \caption{Architecture of the 2FHA protoype}
    \label{fig:my_label}
\end{figure*}

In order to demonstrate the proposed system we created a website that includes a login page and have developed a prototype. The user in order to enter the system  must fill the correct username and password, which is the first authentication factor. Then the system sends to the user a number $M$ that indicates the token that is correct on every login attempt in the future. When logging into the system from a new device, the user must enter the correct OTP. The user receives a number of tokens $N$. He can choose with what platform wants to be alerted for the token, to get it (e-mail, sms, phone call etc.).

\begin{figure}
    \centering
    \includegraphics[width=0.75\linewidth]{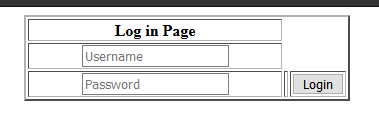}
    \caption{The login page}
    \label{fig:login}
\end{figure}

Then we must enter the second authentication factor. The prototype of the  2FHA mechanism produces 3 qrcodes\cite{qrcodes}, each one of those is represented with a password and sends an sms message\cite{website} to the mobile phone of the user.  The sms includes all 3 OTPs (One Time Password) passwords corresponding to each of the qrcodes \cite{atms}. One is the correct and the others 2 are fake. The user now has to choose what it’s more suitable method for him to continue in order to fill the OTP box and proceed in the website\cite{security}. We ahve to highlight here that the number of produced tokens is kept to 3 only for demonstrating purposes but can be generalized to a number $N$.  

\begin{figure}
\centering
    \includegraphics[width=1.0\linewidth]{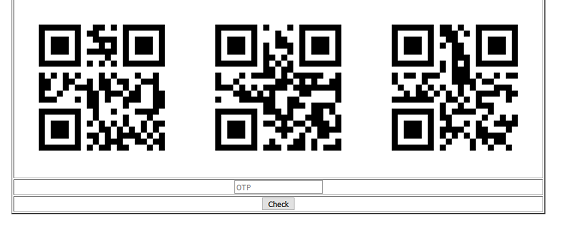}
    \caption{The produced qrcodes}
    \label{fig:qrcodes}
\end{figure}

\begin{figure}
    \centering
    \includegraphics[width=0.95\linewidth]{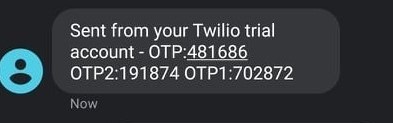}
    \caption{OTP passwords sent as an sms message}
    \label{fig:SMS}
\end{figure}

If the user chooses to scan the qrcodes \cite{QRCODES1}, the process is simple. He scans the correct qrcode and then he fills the OTP box. The qrscanner is free software and most of them are suitable for any device. If the user doesn’t have qrscanner then the option of sms is more convenient for him. The sms message as presented in Figure \ref{fig:SMS}, will be sent to the user the time he logins to the system. As you can see in Figure \ref{fig:SMS} the message contains 3  OTP passwords(OTP, OTP1, OTP2). These are the produced from  the qr codes. Each user knows that only one of the 3 qrcodes is the correct while the other 2 are fake. 

If the user fills the OTP box correctly, he will continue to the system. If not, then he will be sent back to the initial login page and has to follow  the procedure again. Also for precaution reasons the account of the user can be suspended. The OTPs must follow some rules when created; they can't be very similar among them in order to avoid mispelling mistakes.  



\section{Conclusion - Discussion}\label{sec:concl}

In this paper we have taken actions to strengthen the security of a  system against stolen tokens and penetration attempts. The proposed mechanism combines 2FA and Honeyword principles and can be integrated in any existing platform or web application. We plan to improve the system in the future by producing a higher number of qrcodes and passwords that will increase the security. In the prototype of the proposed system  OTP's are sent them through SMS. In the near future we plan to integrate the proposed 2FHA with google and microsoft authenticators.  We also plan to enhance the registration phase in order to make it more secure by encrypting the initial information.

\balance
\bibliographystyle{IEEEtran}
\bibliography{infocom}

\end{document}